\documentclass[aps,
                preprint,
                amsmath,
                amssymb,
                showpacs,
                superscriptaddress]{revtex4-1}
\usepackage{graphicx} 
\usepackage{dcolumn}  
\usepackage{bm}       
\usepackage{amsfonts}
\usepackage{dsfont}

\usepackage{ulem}
\usepackage{color}

\usepackage{ulem}
\usepackage{color}

\begin{document}

\title{Controlling Valley-Polarisation in  Graphene via  Tailored Light Pulses}

\author{M. S. Mrudul}
\affiliation{%
Department of Physics, Indian Institute of Technology Bombay,
            Powai, Mumbai 400076, India }
                      
\author{Gopal Dixit}
\email[]{gdixit@phy.iitb.ac.in}
\affiliation{%
Department of Physics, Indian Institute of Technology Bombay,
            Powai, Mumbai 400076, India }

\date{\today}


\begin{abstract}
Analogous to charge and spin, electrons in solids endows an additional degree of freedom: 
the valley pseudospin.
Two-dimensional hexagonal materials such as graphene exhibit two valleys, labelled as 
$\mathbf{K}$ and $\mathbf{K}^{\prime}$. These  two valleys  have the potential to realise logical operations in two-dimensional materials. Obtaining the desired control over valley polarisation between the two valleys is a prerequisite for the logical operations. 
Recently, it was shown that two counter-rotating circularly polarised laser pulses can induce a significant valley-polarisation in graphene. The main focus of the present work is to optimise the valley polarisation in monolayer graphene by controlling different laser parameters,  such as wavelength, intensity ratio, frequency ratio  and sub-cycle phase in  two counter-rotating circularly polarised laser setup. Moreover, an alternate approach, based on single or few-cycle linearly polarised laser pulse, is also explored to induce significant valley polarisation in graphene. Our work could help 
experimentalists to choose a suitable method with optimised parameter space to obtain  
the desired control over valley polarisation in monolayer graphene. 
\end{abstract}

\maketitle

\section{Introduction}
Graphene is a centrosymmetric two-dimensional material with 
zero bandgap~\cite{novoselov2005two}.  
Massless Dirac equation is used to describe 
the charge carriers with exceptional transport properties, which have lead to the proliferation of 
interesting physical phenomena, such as topological superconductivity or anomalous integer quantum Hall effect~\cite{neto2009electronic, geim2009graphene}.  
One of the most interesting features of graphene  
is that electron possess an extra degree of freedom: the valley pseudospin -- 
Valley is a  minima or maxima in  the conduction or valence bands. 
Graphene has inequivalent and degenerate valleys located at  the corners of the Brillouin zone with 
crystal momenta  $\mathbf{K}$ and $\mathbf{K}^{\prime}$. 
These valleys have  potential  to encode, process and store quantum information~\cite{vitale2018valleytronics}. 

However, despite having all the interesting properties, 
monolayer graphene is not suited for valleytronics as it has zero bandgap 
with zero Berry curvature and exhibits  inversion symmetry~\cite{schaibley2016valleytronics}.  
Successful attempts have been made to break the inversion symmetry, such as  creating 
a heterostructure with hexagonal boron nitride~\cite{gorbachev2014detecting, 
yankowitz2012emergence, hunt2013massive, rycerz2007valley},  
strain and defect  engineering~\cite{grujic2014spin, settnes2016graphene, faria2020valley, xiao2007valley}, which have created  a finite bandgap 
at $\mathbf{K}$ and $\mathbf{K}^{\prime}$ valleys with valley-contrasting Berry curvature.  
This has led to the realisation of valley polarisation in modified monolayer graphene. 

Light is used to realise valley polarisation in the finite bandgap counterpart of graphene such as transition metal dichalcogenides,  as they have valley-contrasting Berry curvature at 
$\mathbf{K}$ and $\mathbf{K}^{\prime}$~\cite{schaibley2016valleytronics}.  
A circularly polarised light, resonant with the 
material's direct bandgap, is used to manipulate  the electronic population at the valleys. 
By choosing the helicity of the light according to 
the optical-valley selection rules, selective excitation at   
$\mathbf{K}$ and $\mathbf{K}^{\prime}$ in gapped-graphene materials is demonstrated~\cite{mak2012control, jones2013optical, gunlycke2011graphene, xiao2012coupled}.  
A pair of non-resonant 
laser pulses were used to excite and control electronic population at desired valleys in 
tungsten diselenide on ultrafast timescale~\cite{langer2018lightwave}.
Recently,  two-color counter-rotating circularly polarized laser pulses 
are employed to break the symmetry between the $\mathbf{K}$ and $\mathbf{K}^{\prime}$  valleys and
induce valley polarisation 
in hexagonal boron nitride and  molybdenum disulfide~\cite{jimenez2019lightwave}. 
Furthermore, by exploiting the carrier-envelope phase (CEP) of short linearly polarised pulse, 
control over valley polarisation in finite bandgap materials is discussed~\cite{jimenez2021sub}. 

Maintaining the homogeneity during the sample preparation of  monolayer transition metal dichalcogenides 
and  hexagonal boron nitride is challenging. 
Moreover, realising experiments on these quantum materials without 
any substrate is not straightforward.  In these regards, monolayer graphene offers  a better alternative
over other analogous gapped-graphene materials. 

Recently it was demonstrated that the significant  valley-polarisation in monolayer  graphene 
can be achieved  using $\omega-2\omega$ bi-circular laser pulses~\cite{mrudul2021light}.
The reason behind the   significant valley-polarisation is attributed to the 
threefold symmetry of the bi-circular laser pulses, which 
matches with the symmetry of the individual valleys of graphene. 
Furthermore, a simple recipe to read out the valley polarisation, using 
high-harmonic generation (HHG) driven by  an additional  $3\omega$ pulse is proposed~\cite{mrudul2021light}. 
In recent years, high-harmonic spectroscopy became a powerful method to probe various aspects of  
electron dynamics in solids~\cite{mrudul2020high, pattanayak2019direct, neufeld2021light, pattanayak2020influence}.  
Also,  the idea of $\omega-2\omega$ bi-circular fields driven HHG  is extended from atomic 
systems~\cite{ansari2021controlling, dixit2018control} 
to solids~\cite{heinrich2021chiral}. 
Moreover, remarkable  works  were   reported on  electron dynamics in graphene via intense laser pulse
~\cite{heide2020sub, kelardeh2016attosecond, heide2018coherent, higuchi2017light},  including Floquet-engineered valleytronics~\cite{kundu2016floquet, friedlan2021valley}.

Present  work is dedicated  to understanding the scalability of 
the valley polarisation in monolayer graphene with respect to 
different laser parameters, such as 
wavelength, intensity, sub-cycle phase of the $\omega-2\omega$ bi-circular laser pulses. 
Other combinations of the tailored pulses like $
\omega-3\omega$ bi-circular laser pulses, and 
 CEP-controlled single or few-cycle linearly polarised pulse will be tested to optimise the valley polarisation. 
The paper is organised as follows:  Theoretical methods are presented in Sec. II, Sec III discusses the results of valley polarisation induced by various tailored laser pulses, conclusion and outlook are presented in Sec. IV. Atomic units are used throughout unless specified otherwise.

\section{Theoretical Methods}

\begin{figure}[h!]
	\includegraphics[width= \linewidth]{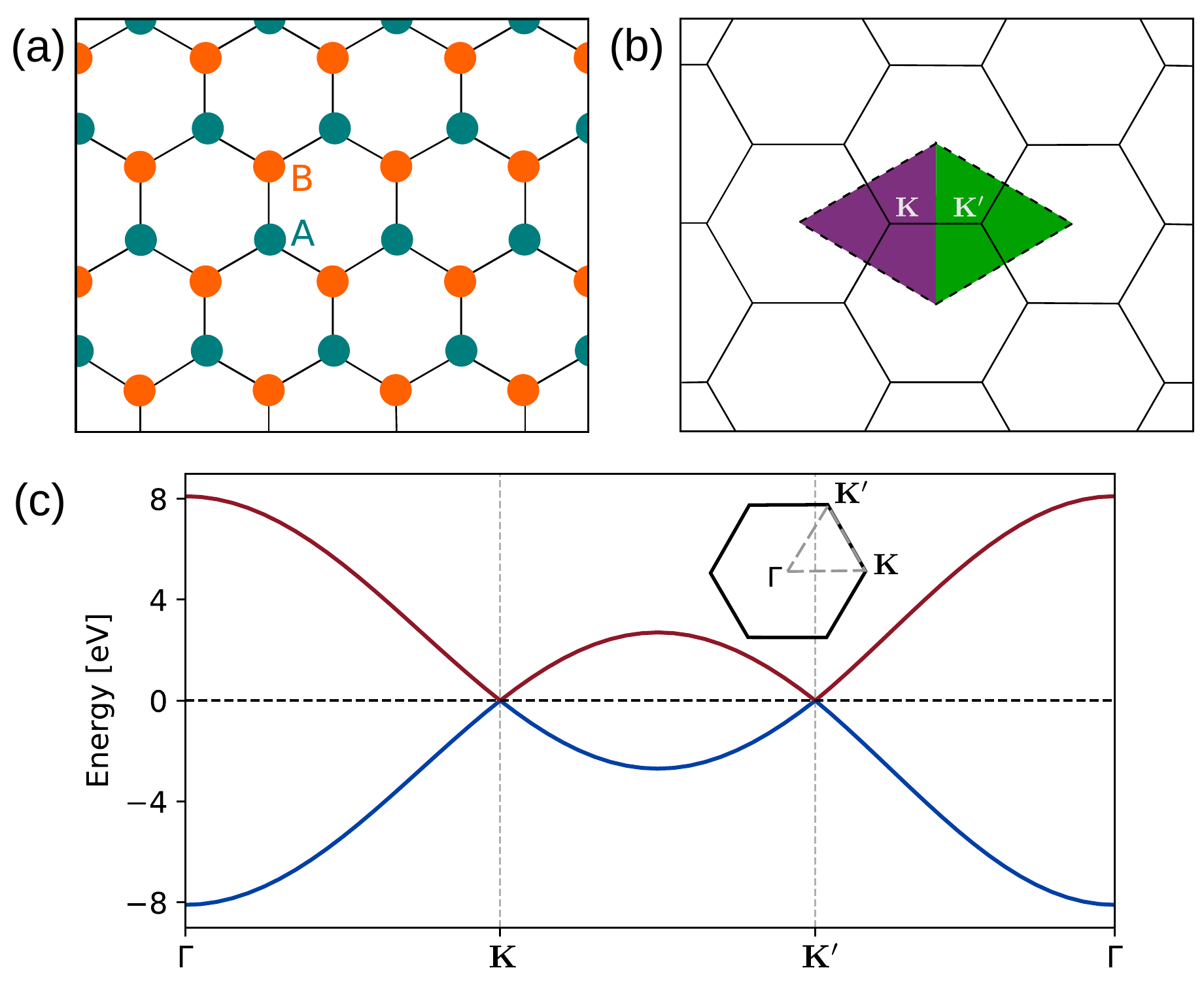}
	\caption{(a)  Honeycomb lattice structure of graphene in real-space. 
	A and B are two inequivalent carbon atoms in the unit cell. (b) The reciprocal-space  lattice structure of graphene. The area within dashed lines is the reciprocal-space unit cell, whereas the areas around $\mathbf{K}$ and $\mathbf{K}^\prime$ valleys for integration are shaded in violet and green, respectively. (c) Energy band-structure of graphenealong high-symmetry directions in the Brillouin zone.}  
	\label{fig1}
\end{figure}

 A real-space lattice structure of 
graphene in which carbon atoms are arranged in a honeycomb lattice is shown in Fig.~\ref{fig1}(a).  
The unit-cell of graphene has two inequivalent carbon atoms marked as A and B in Fig.~\ref{fig1}(a). 
In this case, the lattice parameter is chosen as 2.46~\AA~\cite{reich2002tight}. 
The reciprocal-space lattice of graphene is shown in Fig.~\ref{fig1}(b), where area within the dashed lines is the reciprocal unit-cell. 
In this work, nearest-neighbour tight-binding approximation is considered in which electrons in the p$_z$ orbitals are used to obtain the ground-state of graphene. 
The  Hamiltonian within nearest-neighbour tight-binding approximation is written as 
\begin{equation}
	\hat{\mathcal{H}} = -\gamma f(\textbf{k})\hat{a}^\dagger_\textbf{k}\hat{b}_\textbf{k} + \textrm{H.c.}
	\label{eq1} 
\end{equation}
Here,  the annihilation (creation) operators associated with A and B types of atoms are denoted as 
$\hat{a}_\textbf{k}$ ($\hat{a}^{\dagger}_\textbf{k}$) and $\hat{b}_\textbf{k}$ ($\hat{b}^{\dagger}_\textbf{k}$), respectively. 
$\gamma$ is the nearest neighbour hopping energy with its value  to be 2.7 eV~\cite{reich2002tight,trambly2010localization,moon2012energy}. 
The function $f(\textbf{k})$ is defined as $f(\textbf{k})$ = $\sum_i e^{i\textbf{k} \cdot \delta_i}$ 
with $\delta_i$ as the nearest neighbour vectors of A atom. 
By diagonalising the Hamiltonian in Eq.~(\ref{eq1}), eigenvalues are obtained as 
$\mathcal{E}_{\pm}(\textbf{k}) = \pm \gamma |f(\textbf{k})|$, 
which are plotted along the high-symmetry directions in Fig.~\ref{fig1}(c). 
There are two high-symmetry	points in the reciprocal space, where bandgap vanishes and energy bands have linear dispersion, termed $\mathbf{K}$ and $\mathbf{K}^\prime$ points. 
 These inequivalent high-symmetry  points at the corners of the Brillouin zone are related by time-reversal symmetry.

Time evolution of the density matrix, $\rho_{mn}^{\textbf{k}}$, is performed by solving 
 Semiconductor Bloch equations in Houston basis as~\cite{golde2008high, mrudul2021high}  
\begin{equation}\label{eq2}
\begin{split}
\frac{\partial}{\partial t}  \rho _{mn}^{\textbf{k}}  =& -i \mathcal{E}_{mn}^{\textbf{k}+\textbf{A}(t)}\rho_{mn}^{\textbf{k}} -(1-\delta_{mn})\frac{\rho_{mn}^{\textbf{k}}}{T_2}\\  &+ i \textbf{F}(t) \cdot \sum_l \left(\textbf{d}_{ml}^{~\textbf{k}+\textbf{A}(t)}\rho_{ln}^{\textbf{k}} - \textbf{d}_{ln}^{~\textbf{k}+\textbf{A}(t)}\rho_{ml}^{\textbf{k}}  \right).
\end{split}
\end{equation}
Here,  $\mathcal{E}^{\textbf{k}}_{mn}$ and  \textbf{d}$_{mn}^\textbf{k}$ are, respectively, 
energy-gaps and dipole-matrix elements between $|m,\textbf{k}\rangle$ and $|n,\textbf{k}\rangle$ 
states in the Brillouin zone. 
\textbf{d}$_{mn}^\textbf{k}$ is defined as \textbf{d}$_{mn}^\textbf{k}$ = -i$\left\langle u_m^{\textbf{k}} |\nabla_{\textbf{k}}| u_n^\textbf{k}\right\rangle$, where $u_n^\textbf{k}$ 
is the periodic part of the Bloch function.
$\textbf{F}(t)$ and $\textbf{A}(t)$ are, respectively, the electric field and vector potential associated with the laser pulse and 
are related as $\textbf{F}(t)$ = -${\rm d} \textbf{A}(t)/{\rm d}t$. 
A phenomenological term accounting for the decoherence between electron and hole is 
included with a constant dephasing time $T_2$. The term accounting for the 
population relaxation, $T_1$, is neglected assuming $T_1\gg T_2$.
The coupled differential equations in Eq.~(\ref{eq2}) are numerically solved using the fourth-order Runge-Kutta method with a time-step of 0.02 fs. The reciprocal space is sampled with a 180$\times$180 grid. The value of the dephasing time $T_2$  is chosen as  10 fs.

We define the valley-asymmetry parameter to quantify valley polarization as
\begin{equation}\label{eq3}
\eta = \frac{n_{c}^{\mathbf{K}^{\prime}} - n_{c}^{\mathbf{K}}}{(n_{c}^{\mathbf{K}^{\prime}} + n_{c}^{\mathbf{K}})/2},
\end{equation}
where $n_{c}^{\mathbf{K}}$ and $n_{c}^{\mathbf{K}^\prime}$ are the residual conduction band electron populations around $\mathbf{K}$ and $\mathbf{K}^\prime$ valleys, respectively. 
We estimate the total residual conduction band population, $n_c$, by integrating $\rho_{cc}^\textbf{k}$ in the reciprocal-space unit cell [see Fig.~\ref{fig1}(b)] at the end of the laser pulse. 
$n_c^{\mathbf{K}}$ and $n_c^{\mathbf{K}^\prime}$ are obtained by integrating, respectively, 
within the violet and green shaded regions in Fig.~\ref{fig1}(b), such that $n_c = n_c^{\mathbf{K}} + n_c^{\mathbf{K}^\prime}$. 

The tailored laser field is a superposition of two counter-rotating circularly polarized pulses  
with photon energies $\omega_1=\omega$ and $\omega_2=n\omega$, respectively.  
This is known as $\omega-n\omega$ bi-circular field. 
The vector potential corresponds to this tailored field is defined as
\begin{equation}\label{eq4}
\begin{split}
	\textbf{A}(t) = \frac{A_0 f(t)}{\sqrt{2}} \bigg(&\left[ \cos(\omega t + \phi) + \frac{\mathcal{R}}{n} \cos(n\omega t)\right] \hat{\textbf{e}}_x
	+\\ &\left[\sin(\omega t + \phi) - \frac{\mathcal{R}}{n} \sin(n\omega t)\right]\hat{\textbf{e}}_y\bigg).
\end{split}
\end{equation}
Here, $A_0=F_{\omega}/\omega$ is the amplitude of the vector potential
of the fundamental laser for which $F_{\omega}$ is the strength of the electric field. $f(t)$ is the 
temporal envelope of the driving field. The two laser fields have a sub-cycle phase difference of $\phi$ 
and the ratio between the two electric-field strengths  is denoted by $\mathcal{R}$.  
In the following section, 
we will discuss results obtained for a seven-cycle pulse with a sin-squared envelope as 
used in Ref.~\cite{mrudul2021light}. 

\section{Results and Discussion}

\begin{figure}[h!]
	\includegraphics[width= \linewidth]{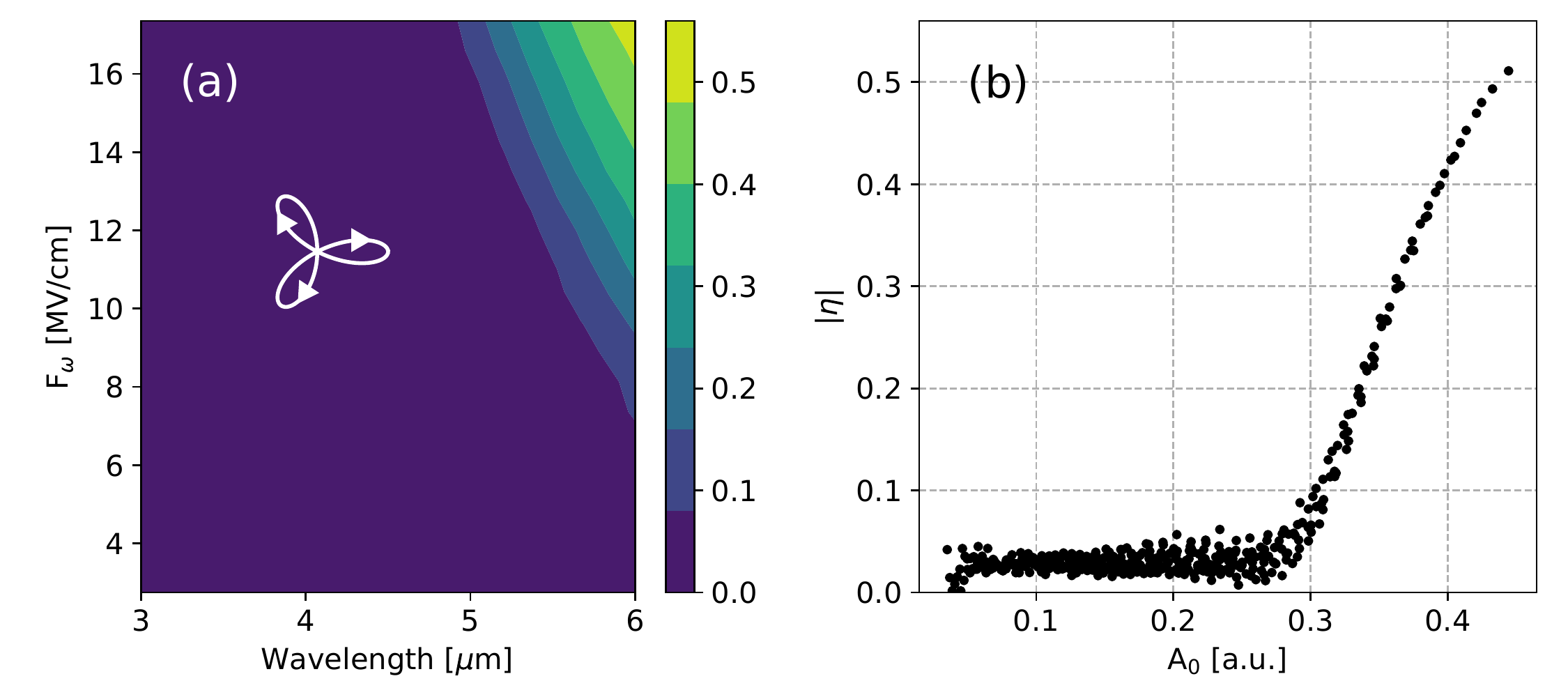}
	\caption{Dependence of the valley polarisation on  the laser parameters. (a) Valley-asymmetry 
	parameter ($\eta$) as a function of wavelength and electric field amplitude of the fundamental 
	pulse.  (b) $\eta$ as a function of the amplitude of the vector potential 
	(A$_0$ = F$_\omega/\omega$). Lissajous figure of the vector potential is shown in the inset 
	of (a).} 
	\label{fig2}
\end{figure}

Figure~\ref{fig2}(a) presents the variation in the valley polarisation as
a function of electric field strength ($F_\omega$) and  wavelength ($\lambda$). 
$F_\omega$ and $\lambda$ are varied in the range 3-17 MV/cm and 3-6 $\mu$m, respectively. 
The value of the sub-cycle phase  ($\phi$) is chosen as 0$^{\circ}$
for which the valley polarisation is maximum. 
As evident from the figure, $\eta$ value corresponding to 
longer wavelengths and intense pulses is higher than 50\%. 
We observe that  there is no considerable valley polarisation up to a threshold value of $F_\omega$ and $\lambda$. Once the threshold values are reached, 
$\eta$ increases monotonically as a function of both $F_\omega$ and $\lambda$. 

To have a better understanding, the data of Fig.~\ref{fig2}(a) is represented 
as a function of $A_0$ of the $\omega$-field in Fig.~\ref{fig2}(b). 
It is apparent that $\eta$ increases linearly with respect to $A_0$, after reaching the threshold value, 
as reflected from Fig.~\ref{fig2}(b).
This findings  can be directly correlated to the mechanism of valley-polarisation in monolayer 
graphene as discussed in Ref.~\cite{mrudul2021light}.
The electron dynamics in $\mathbf{K}$ and $\mathbf{K}^\prime$ valleys acts differently when the electrons are driven out of the isotropic part of the valleys in the reciprocal space using 
$\omega-2\omega$ bi-circular field. This mechanism causes different population buildup near the two valleys, resulting in considerable valley-polarisation.
It is known that the 
dynamics of electron's crystal momentum follows the vector potential of the electron in the reciprocal space. 
Therefore, the excursion length of the electron in the reciprocal space increases as the strength of the vector potential increases. 
It is evident that we don't see any significant valley-polarisation up to the threshold value of the vector potential as the electron in the conduction band, generated  close to the $\mathbf{K}$ points, 
exhibiting dynamics still in the isotropic part of the energy landscape.
Once the electron reaches to the anisotropic part, the valley-polarisation scales linearly 
with $A_0$ as expected. 
It is important to mention that a further increase  in the field strength or wavelength can result in mixing of electron population of two valleys, and the meaning of valley polarisation becomes ill-defined.  

\begin{figure}[]
\includegraphics[width=\linewidth]{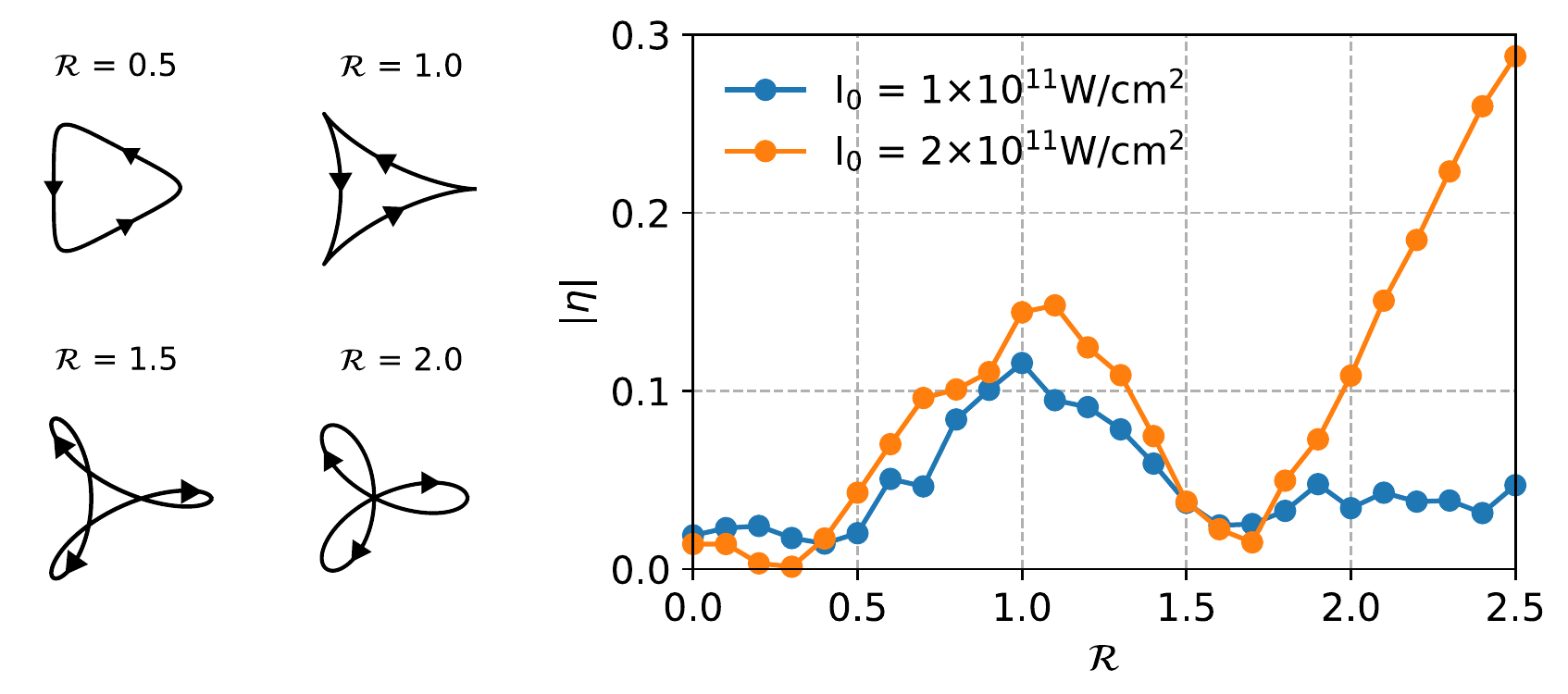}
\caption{ Effect of the ratio of the electric-field strengths, $\mathcal{R}$, on the valley polarisation.  
 Left panel: Lissajous figures of vector potential for different ratios of the field strengths of the 
 two pulses ($\mathcal{R}$). Right panel: Valley asymmetry parameter, $\eta$, 
 as a function of $\mathcal{R}$ for fundamental laser intensities (I$_0$) of  1$\times$10$^{11}$ W/cm$^2$ (blue, F$_\omega$ $\approx$ 9 MV/cm ) and 2$\times$10$^{11}$ W/cm$^2$ (orange, F$_\omega$ $\approx$ 12 MV/cm).  Here, laser pulses of 6 $\mu$m wavelength are used.} 
\label{fig3}
\end{figure}

Now let us explore how the valley polarisation changes by changing  the relative 
strength of the electric fields ($\mathcal{R}$) in $\omega-2\omega$ configuration. 
By varying the value of $\mathcal{R}$, Lissajous figure of the total vector potential 
changes substantially, while the three-fold symmetry is preserved  
as shown in the left panel of the Fig.~\ref{fig3}.
We present $\eta$ as a function of $\mathcal{R}$ for laser intensities of 1$\times$10$^{11}$W/cm$^2$ (blue, F$_\omega$ $\approx$ 9 MV/cm) and  2$\times$10$^{11}$W/cm$^2$ (orange, F$_\omega$ $\approx$ 12 MV/cm). 
As evident from the figure, the valley polarisation is maximum for $\mathcal{R}$ = 1. 
Moreover, there is a linear increase in the valley polarisation after $\mathcal{R}$ = 1.7 for  
laser intensity of 2$\times$10$^{11}$ W/cm$^2$.  

Let us recall  that the contrasting  dynamics of electron in two valleys is attributed 
to the fact that the symmetry of the vector potential matches with one of the valley, and not with the other. 
If this is the case then the valley polarisation  
will be most efficient when the total field resembles close to the energy landscape of one of the 
valleys for which the symmetry matches. 
This is what happens here, close to $\mathcal{R}$ = 1,  
the three-fold symmetry of the $\omega-2\omega$ fields 
matches closer to the energy-landscape of the conduction band, resulting in a higher valley polarisation. 
On the other hand, linear increase of the valley polarisation for 2$\times$10$^{11}$ W/cm$^2$ is due 
to increase of the peak of the resultant electric field with $\mathcal{R}$. 
The maximum field-strength scales with $\mathcal{R}$ as F$_{max}$ = (1+$\mathcal{R}$)F$_\omega$. 
Therefore, a linear increase in the resultant field strength for a particular value of $\mathcal{R}$ = 1.7
makes the field strong enough to push the electron out of the isotropic part. 

\begin{figure}[h!]
	\includegraphics[width=\linewidth]{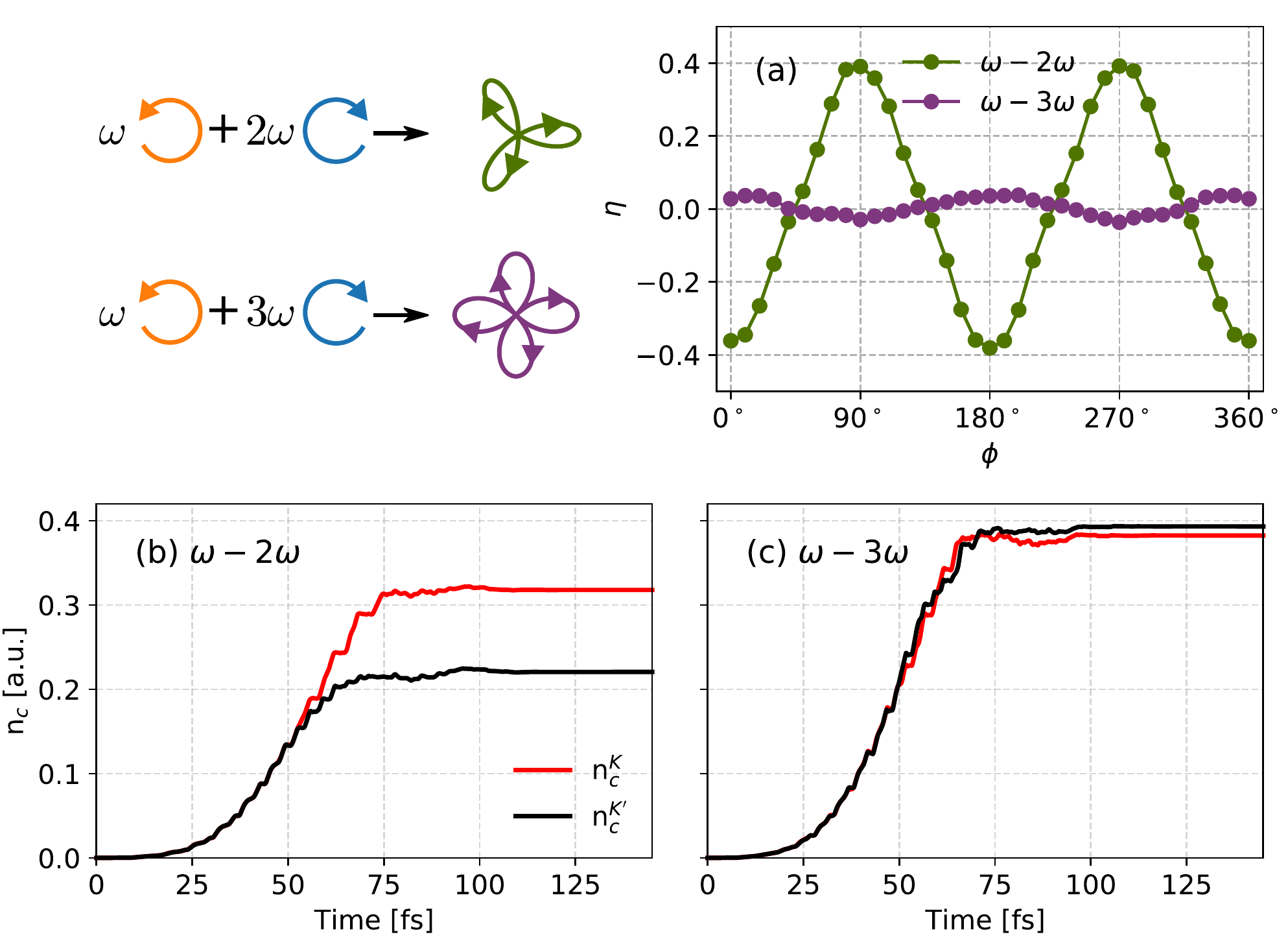}
	\caption{Valley polarisation induced by different tailored fields. (a) Valley polarisation in 
	$\omega-2\omega$ (green) and $\omega-3\omega$ (violet) bi-circular counter-rotating fields  
	as a function of sub-cycle phase ($\phi$).  Lissajous figures corresponding to $\omega-2\omega$
	and $\omega-3\omega$ fields are shown in the top left panel. 
	Conduction band populations around $\mathbf{K}$ (red) and $\mathbf{K}^\prime$ (black) valleys 
	during the laser pulse for (b) $\omega-2\omega$ and (c) $\omega-3\omega$  bi-circular 
	counter-rotating fields.} 
	\label{fig4}
\end{figure}

So far we have investigated how different parameters of the $\omega-2\omega$ bi-circular field affect the valley polarisation.  Let us ask an interesting  question: is it only possible to 
achieve such a high-degree of valley polarisation using $\omega-2\omega$ bi-circular pulses or 
there are other configurations of tailored pulses for that purpose.  
To answer this question, we consider 
$\omega-3\omega$ bi-circular pulses and investigate 
what amount of valley polarisation is achievable. 
We use a laser intensity of 3$\times$10$^{11}$W/cm$^2$ and a wavelength of 6 $\mu$m. 
Moreover, $\mathcal{R} = 2$ and $\mathcal{R} = 3$  are  used for $\omega-2\omega$ and $\omega-3\omega$, respectively.  
Lissajous figures of the total vector potentials  for both the configurations are presented in Fig.~\ref{fig4}.  As evident from the figure, $\omega-2\omega$ and $\omega-3\omega$ fields have three-fold and four-fold symmetries, respectively.  

The valley polarisation as a function of $\phi$  
for $\omega-2\omega$ and $\omega-3\omega$ fields are presented  in Fig.~\ref{fig4}(a). 
As discussed earlier, $\omega-2\omega$ field generates considerable valley polarisation,  
and is modulated as a function of $\phi$. In contrast,  $\omega-3\omega$ field generates feeble valley polarisation. 
We have also performed the same comparison for a laser intensity of 1$\times$10$^{11}$W/cm$^2$ and obtained similar observations (not shown here). 

The conduction band population around $\mathbf{K}$ and $\mathbf{K}^\prime$ valleys during the laser pulse for  $\omega-2\omega$ and $\omega-3\omega$ fields  are 
shown in Fig.~\ref{fig4}(b) and (c), respectively. 
The total conduction band population is higher in the case of $\omega-3\omega$ field, owing to the relatively higher strength of the field. 
In the case of $\omega-3\omega$ field, valley-polarisation fluctuates between two valleys during the laser propagation   without attaining significant  valley polarisation. On the other hand, a considerable, and consistent 
valley polarisation is visible for $\omega-2\omega$ field, once the  threshold field strength is reached. 
In short, it is essential to use a field which breaks the inversion symmetry of the monolayer 
graphene to achieve a significant valley polarisation. 
Now let us explore an alternate way  to  break the inversion symmetry  
by using single or few-cycle pulses and analyse the valley-polarisation generated.

\begin{figure}[h!]
	\includegraphics[width=\linewidth]{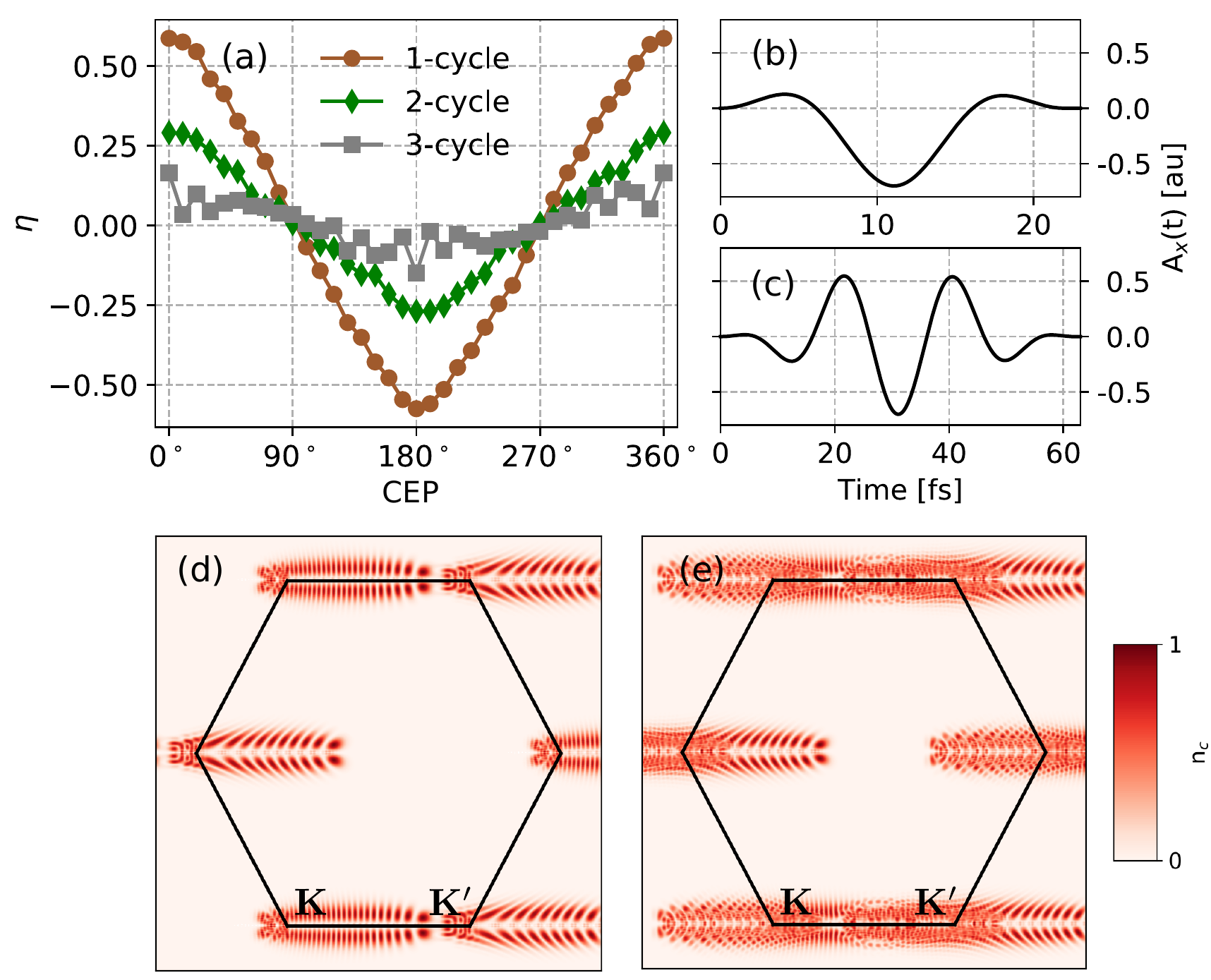}
	\caption{Valley polarisation for a  single/few-cycle  linearly polarised pulse. 
	(a) $\eta$ as a function of carrier envelope phase (CEP) 
	for a single-cycle (brown), two-cycle (green), 
	and three-cycle linearly polarised pulses. 
	Vector potentials for single-cycle (b) and three-cycle (c) pulses. 
	Residual conduction band population in the first Brillouin zone induced by 
	(d) single-cycle and (e) three-cycle pulses.} 
	\label{fig5}
\end{figure}

Recently, single and two-cycle circularly polarised pulses have been employed to achieve 40$\%$-60$\%$ valley polarisation in  transition metal dichalcogenides.  It is found that the 
right-handed pulse prefer to populate $\mathbf{K}$-valley, whereas left-handed pulse prefer 
$\mathbf{K}^{\prime}$-valley~\cite{motlagh2018}. This effect is due to the topological resonance caused by laser-driven dynamics of Bloch electrons~\cite{motlagh2018}.
Lately, it was demonstrated that an ultrashort few-cycle 
linearly polarised pulse along $\Gamma-\mathbf{K}$ direction can induce significant valley polarisation 
in gapped-graphene materials, such as hexagonal boron nitride and transition metal dichalcogenides~\cite{jimenez2021sub}. 

Let us revisit this mechanism of valley polarisation using short, linearly polarised pulse in gapped-graphene hexagonal materials as follows: The photon energy of the laser pulse is chosen to be much smaller than the band-gap of the material. In this case, electrons are transferred to the conduction band only close to the peak of the electric field.
In contrast to relatively long linearly polarised pulses, the maximum electric field doesn't imply zero vector potential for single or few-cycle pulses.    
Exploiting this fact, the electrons injected at the peak of the electric field is traversed in the 
momentum-space depending on the value of vector potential at that time. 
This mechanism results in a finite 
conduction band population transfer to one of the valleys of the material. The direction of electron dynamics can be controlled by changing the CEP of the pulse.
This approach works only when the polarisation of the laser is along $\Gamma-\mathbf{K}$ direction
as the mechanism of the valley polarisation requires the laser to have a direction connecting $\mathbf{K}$ and $\mathbf{K}^\prime$. 
As this method doesn't depend upon the material's Berry curvature, it applies to centrosymmetric materials. So it was anticipated that the same technique could be employed to graphene -- a zero band-gap material~\cite{jimenez2021sub}.

Figure~\ref{fig5}(a) present $\eta$ as a function of CEP 
for a single-cycle (brown), two-cycle (green), and three-cycle (grey) linearly  polarised pulses 
along the $\Gamma-\mathbf{K}$ direction. 
The pulse has a peak intensity of 10$^{12}$ W/cm$^2$ and a wavelength of 6 $\mu$m. 
It is evident from the figure that a single-cycle pulse can induce 
the valley polarisation above 50$\%$. 
However, the valley polarisation decreases drastically as the number of cycle of the pulse increases, which can be explained as: Electrons can be injected to conduction band at any time during laser propagation. 
They can be traversed in conduction band as a function of the vector potential. 
If the laser pulse allows population dynamics preferably in +ve or -ve direction from $\mathbf{K}$-point  
then one of the valleys will be populated. 
As the pulse start becoming long by increasing the number of cycles, the preferred direction of the polarisation diminishes.
As a result, the valley polarisation reduces significantly for long linearly polarised pulses  as reflected from Fig.~\ref{fig5}(a).
Note that, in contrast to gapped-graphene situation, 
electron transfer in monolayer graphene is not limited to close to the peak of the laser pulse. 

The strength of the valley polarisation modulates as a function of the pulse's CEP as  apparent 
from the figure. 
The vector potential corresponding to single-cycle, and three-cycle pulses are shown 
in Figs.~\ref{fig5}(b) and (c), respectively. 
The variation in  $\eta$  as a function of CEP can be understood as follows: 
The pulse is asymmetric in time for CEP = 0$^{\circ}$, which translate to the transfer of electron from  one valley is preferable over the other valley. 
As the CEP changes from 0$^{\circ}$ to 180$^{\circ}$, preference  
changes to other valley over the previous one. 
The vector potential  becomes symmetric at CEP values of 90$^{\circ}$ and 270$^{\circ}$, which 
leads to no valley polarisation.

The residual conduction band population corresponding to the vector potentials in Figs.~\ref{fig5}(d) and (e) are presented  in Figs.~\ref{fig5}(b) and (c), respectively. 
It can be observed from Fig.~\ref{fig5}(b) that a short single-cycle pulse can push electrons towards a particular direction with respect to the $\mathbf{K}$-point, whereas three-cycle (long) pulse 
results in uniform distribution of conduction band population with respect to the polarisation axis. 
The conduction band population in Fig.~\ref{fig5}(e) shows interference of the electronic population from two valleys since electrons are moved in both directions in this case and reduces the value of $\eta$. 
These interfering electrons of two valleys also result in the noisy structure of $\eta$ as reflected in  Fig.~\ref{fig5}(a).

\section{Conclusions and Outlook}
In summary, we have investigated few scenarios of optimising the valley polarisation in monolayer 
graphene using different kinds of tailored light pulses. In case of $\omega-2\omega$ bi-circular tailored 
pulses, the valley polarisation increases with the wavelength of the pulse. Moreover,    
the valley polarisation scales linearly with the strength of the vector potential. 
However, the value of the vector potential can't be increased arbitrary as the electronic populations from both the valleys start overlapping and the definition of the valley polarisation becomes ill-defined. Also,  intensity ratio of the $\omega-2\omega$ pulses plays an important role in 
the valley polarisation, and it is optimum for unit ratio. 
The sub-cycle phase of the $\omega-2\omega$ pulses provides another control knob, and 
the valley polarisation can be modulated over two valleys by tuning the sub-cycle phase. 
It has been found that the valley polarisation reduces drastically 
if one frequency in the bi-circular setup is changed  from 
$\omega-2\omega$ to $\omega-3\omega$.
Also, our findings are robust against a dephasing time of 30 fs, i.e., interband decoherence time between electron and hole.  

Single and few-cycle  linearly  polarised pulses are also tested to induce  valley polarisation. 
It has been observed that the single-cycle pulse can induce valley polarisation of similar strength as 
observed in the case of the $\omega-2\omega$ pulses. As the number of cycles in the pulse  
increases, valley polarisation reduces significantly. 
In this case, CEP provides a control knob to  modulate the valley polarisation from one valley to other. 
The underlying mechanisms of valley polarisation in $\omega-2\omega$ bi-circular 
and single or few-cycle  linearly  polarised pulses are entirely different. 
However, while fixing the laser parameters for a particular method, one needs to  
take the damage threshold of graphene into considereation~\cite{roberts2011response}.
Furthermore, high-harmonic spectroscopy would be an appropriate appraoch to read the valley polarisation induced by both the methods~\cite{jimenez2021sub, mrudul2021light}.   

We think that controlling the sub-cycle phase between $\omega-2\omega$ bi-circular long pulses is ``relatively easy" in comparison to the generation and control of a single-cycle pulse. 
Moreover, 
one could imagine  of employing half-cycle 
or fraction of single-cycle pulse to improve the valley contrast, but  it will be a daunting task experimentally.  It will be interesting to see the realisation of different  logical operations and device fabrications using the valley polarisation in monolayer graphene as it has been done in similar materials~\cite{jana2021robust, sin2017valleytronics}. 
Moreover, the bandgap at $\mathbf{K}$ and $\mathbf{K}^\prime$ remain zero as we move monolayer to bilayer graphene. Also, the energy contour in bilayer graphene exhibits trigonal structures around $\mathbf{K}$ and $\mathbf{K}^\prime$ valleys. Therefore, we believe that our findings remain valid for bilayer graphene qualitatively.

\section*{Acknowledgments}
G. D. acknowledges support from Science and Engineering Research Board (SERB) India 
(Project No. ECR/2017/001460) and the Ramanujan fellowship (SB/S2/ RJN-152/2015). 
G. D. acknowledges fruitful discussion with Prof. Misha Ivanov.


\end{document}